\documentclass[conference]{IEEEtran}
\IEEEoverridecommandlockouts
\pdfoutput=1
\usepackage{cite}
\usepackage[absolute,showboxes]{textpos}
\usepackage{amsmath,amssymb,amsfonts}
\usepackage{algorithmic}
\usepackage{graphicx}
\usepackage{textcomp}
\usepackage{xcolor}
\usepackage{todonotes}
\usepackage{nohyperref} 
\usepackage[binary-units,per-mode=symbol]{siunitx}
\usepackage{subcaption}
\usepackage[nohyperlinks,nolist]{acronym}
\usepackage{tikz}
\usetikzlibrary{patterns}
\usepackage{pgfplots}
\pgfplotsset{compat=newest}
\usepgfplotslibrary{units}
\makeatletter
\pgfplotsset{
 unit code/.code 2 args=
   \begingroup
   \protected@edef\x{\endgroup\si{#2}}\x
}
\makeatother
\usetikzlibrary{pgfplots.groupplots}
\definecolor{lightcorered}{RGB}{250,150,150}
\definecolor{corered}{RGB}{200,0,0}
\definecolor{coredarkgray}{RGB}{51,51,51}
\definecolor{coreblue}{RGB}{0, 46, 125}

\begin{document}

\title{Software-Defined Networks Supporting Time-Sensitive In-Vehicular Communication}

\author{\IEEEauthorblockN{Timo H\"ackel, Philipp Meyer, Franz Korf and Thomas C. Schmidt}
\IEEEauthorblockA{\href{http://www.haw-hamburg.de/ti-i}{\textit{Dept. Computer Science}},
\href{http://www.haw-hamburg.de/ti-i}{\textit{Hamburg University of Applied Sciences}}, Germany \\
\{\href{mailto:timo.haeckel@haw-hamburg.de}{timo.haeckel}, \href{mailto:philipp.meyer@haw-hamburg.de}{philipp.meyer}, \href{mailto:franz.korf@haw-hamburg.de}{franz.korf}, \href{mailto:t.schmidt@haw-hamburg.de}{t.schmidt}\}@haw-hamburg.de
}}

\maketitle

\setlength{\TPHorizModule}{\paperwidth}
\setlength{\TPVertModule}{\paperheight}
\TPMargin{5pt}
\begin{textblock}{0.8}(0.1,0.02)
     \noindent
     \footnotesize
     If you cite this paper, please use the VTC reference:
     T. H{\"a}ckel, P. Meyer, F. Korf and T.~C. Schmidt. Software-Defined Networks Supporting Time-Sensitive In-Vehicular Communication. In \emph{Proc. of IEEE VTC2019-Spring}, IEEE, 2019.
\end{textblock}

\begin{abstract}
Future in-vehicular networks will be based on Ethernet. 
The IEEE Time-Sensitive Networking (TSN) is a promising candidate to satisfy real-time requirements in future car communication. 
Software-Defined Networking (SDN) extends the Ethernet control plane with a programming option that can add much value to the resilience, security, and adaptivity of the automotive environment.
In this work, we derive a first concept for combining Software-Defined Networking with Time-Sensitive Networking along with an initial evaluation. 
Our measurements are performed via a simulation that investigates whether an SDN architecture is suitable for time-critical applications in the car. 
Our findings indicate that the control overhead of SDN can be added without a delay penalty for the TSN traffic when protocols are mapped properly. 
\end{abstract}


\begin{acronym}
	\acro{AVB}[AVB]{Audio Video Bridging}
	\acro{ARP}[ARP]{Address Resolution Protocol}
	\acro{BE}[BE]{Best-Effort}
	\acro{CAN}[CAN]{Controller Area Network}
	\acro{CBM}[CBM]{Credit Based Metering}
	\acro{CBS}[CBS]{Credit Based Shaping}
	\acro{CMI}[CMI]{Class Measurement Interval}
	\acro{CoRE}[CoRE]{Communication over Realtime Ethernet}
	\acro{CT}[CT]{Cross Traffic}
	\acro{DoS}[DoS]{Denial of Service}
	\acro{DPI}[DPI]{Deep Packet Inspection}
	\acro{ECU}[ECU]{Electronic Control Unit}
	\acroplural{ECU}[ECUs]{Electronic Control Units}
	\acro{IA}[IA]{Industrial Automation}
	\acro{IEEE}[IEEE]{Institute of Electrical and Electronics Engineers}
	\acro{IoT}[IoT]{Internet of Things}
	\acro{IP}[IP]{Internet Protocol}
	\acro{ICT}[ICT]{Information and Communication Technology}
	\acro{LIN}[LIN]{Local Interconnect Network}
	\acro{MOST}[MOST]{Media Oriented System Transport}
	\acro{OEM}[OEM]{Original Equipment Manufacturer}
	\acro{RC}[RC]{Rate-Constrained}
	\acro{SDN}[SDN]{Software-Defined Networking}
	\acro{SR}[SR]{Stream Reservation}
	\acro{SRP}[SRP]{Stream Reservation Protocol}
	\acro{TCP}[TCP]{Transmission Control Protocol}
	\acro{TDMA}[TDMA]{Time Division Multiple Access}
	\acro{TSN}[TSN]{Time-Sensitive Networking}
	\acro{TSSDN}[TSSDN]{Time-Sensitive Software-Defined Networking}
	\acro{TT}[TT]{Time-Triggered}
	\acro{TTE}[TTE]{Time-Triggered Ethernet}
	\acro{UDP}[UDP]{User Datagram Protocol}
	\acro{QoS}[QoS]{Quality-of-Service}
\end{acronym}


\section{Introduction} 
\label{sec:introduction}
Over the past years, Ethernet has emerged as the next high bandwidth communication technology for in-car networks\cite{mk-ae-15}. 
There were several attempts to introduce support for real-time requirements of which the IEEE collection of standards for \ac{TSN}\cite{ieee8021q-18} is the most promising. 
The use of Ethernet in automotive networks enables the adoption of multiple standards from the internet domain, such as the internet protocols and transport protocols. 
\\
An emerging trend in Ethernet networks is \ac{SDN}. 
\ac{SDN} has proven especially useful in well-known local area networks such as data centers, as it decreases the complexity and management effort\cite{kfrvr-sdncs-14}. 
The basic approach of \ac{SDN} is to replace the switches with simple forwarding devices and connect them to a logically centralised intelligent network controller \cite{mabpp-oeiJR-08}. 
This allows the execution of complex flow-based forwarding rules in simple and inexpensive devices. 
In the automotive environment \ac{SDN} could provide benefits regarding safety, robustness, security, cost efficiency, and future-readiness with easily updatable network devices. 

In this paper, we analyse the integration of \ac{TSN} and \ac{SDN} to \ac{TSSDN} for in-vehicular networks and explore its potentials and expectations. 
We contribute a mapping for deploying time-sensitive traffic in \ac{TSSDN} switches and describe our methodology of registering time-sensitive flows with OpenFlow. 
We implement and evaluate this architecture in OMNeT++, a common event-based network simulation environment and conduct first evaluations on the real-time capabilities. 
Overall, we show that time-sensitive traffic performance remains unaltered when adding SDN control logic.

This paper is structured as follows.
Section~\ref{sec:background_&_related_work} provides background knowledge and gives an overview on related work. 
The potentials and expectations of \ac{TSSDN} are analysed in Section~\ref{sec:potential_&_expectations}. 
Section~\ref{sec:architecture_proposal} describes the evolution of the switch architecture to \ac{TSSDN} switches and how time-sensitive flows can be implemented with OpenFlow.
In Section~\ref{sec:case_study} a case study is conducted showing the real-time capability of the presented \ac{TSSDN} concept. 
Finally, Section~\ref{sec:conclusion_&_future_work} concludes this work and gives an outlook on future work.


\section{Background \& Related Work} 
\label{sec:background_&_related_work}
\label{sub:time_sensitive_networking}
\label{sub:software_definied_networking}

\ac{TSN} is a set of standards which are defined by the \ac{TSN} task group \cite{ieee8021tsn} of the IEEE. 
IEEE 802.1Q-2018 \cite{ieee8021q-18} extends Ethernet with the ability to forward concurrent real-time- and cross-traffic. 
For supporting a wide range of \ac{QoS} requirements, \ac{TSN} supports several real-time traffic classes. 
These can be synchronous (\ac{TDMA}) or asynchronous such as \ac{TSN}s predecessor \ac{AVB}, which we analysed in former work \cite{slksh-tiice-12}. 
The draft IEEE P802.1Qcc is an amendment to the \ac{TSN} standards and provides enhancements to the \ac{SRP}. 
Besides performance improvements, the draft introduces a controller for central network management. 
However, this controller is only a centralised configuration unit. 
It neither specifies a vendor neutral standardised interface between the controller and the switches, nor extracts the control plane functionality of the network devices.

\ac{SDN} enables a standardised configuration of forwarding devices by an OpenFlow controller, by separating the control and data plane of the network devices~\cite{mabpp-oeiJR-08}.
Kreutz et al. describe the paradigms and concepts of \ac{SDN} in their comprehensive survey~\cite{kfrvr-sdncs-14}. 
The network logic is split into three layers: 
(1) The data plane on which each switch forwards packets according to flow rules, 
(2) the control plane on which each switch is connected to a logically (not necessarily physically) centralised controller that manages the forwarding logic, and 
(3) the management plane on which network administrators manage the controller applications. 
The communication between the \ac{SDN} controller and the switches is specified in the OpenFlow standard of the ONF \cite{onfts025-15}. 

Thiele \& Ernst \cite{te-fabJR-16} show by formal analysis that the concept of \ac{SDN} is generally suitable for real-time environments, especially if the flows are implemented in all switches prior to the data exchange. 
They derive possible worst case boundaries for network configuration latency and deem \ac{SDN} applicable for admission control and fault recovery in automotive networks.
\\
As the network requirements of the \ac{IA} are similar to in-car networks, research in this area can often be transferred directly to vehicles. 
First efforts to create a real-time capable \ac{SDN} have been taken in the \ac{IA}.
Herlich et al. present the idea of a real-time Ethernet \ac{SDN} \cite{hdsd-pcsJR-16} and show typical use cases for \ac{IA} networks as a proof of concept. 
Some of these use cases are directly related to in-vehicular networks, for instance supporting arbitrary network topologies, central and dynamic network (re-)configurations, and fast fail-over mechanisms at network level. 
\\
Nayak et al. \cite{ndr-sdeJR-15} contribute research regarding robustness and reconfiguration in \ac{IA} networks by using exclusive links for real-time traffic. 
They go one step further in developing a Time-Sensitive Software-Defined Network with a scheduling process for time-triggered traffic \cite{ndr-tssJR-16}. 
However, the schedule is not programmed into the switches, but instead the hosts in the network know the full schedule and send data in their time slots. 
The switches are not real-time capable and unscheduled cross traffic could change the network behaviour.
\\
Kobzan et al. share their concepts of software-defined networks for the production plants of the future in the FlexSi-Pro research project\cite{ksabo-stsJR-18}. They evaluate the combination of \ac{TSN} and \ac{SDN} for \ac{IA} and declare the central configuration of real-time traffic with \ac{SDN} as an open research task.

Fussey and Parisis summarised the advantages and disadvantages of using \ac{SDN} in vehicles focusing on enabled features \cite{fp-pvsJR-17}. 
We expand this list focusing on real-time capabilities and benefits that \ac{SDN} can have on \ac{TSN}.
\\
Halba et al. use \ac{SDN} for data interoperability and robustness in the in-vehicular network and thus, substantiate the benefit of \ac{SDN} for in-car networks. 
They show \ac{SDN} controller applications implementing robustness and safety with fast fail-over mechanisms \cite{hmg-rsaJR-18}. 
However, they do not consider real-time requirements in their network design.

As shown above, there were already some efforts in applying \ac{SDN} to automotive networks and making them real-time capable. 
However, the combination of the \ac{TSN} standards with \ac{SDN} remains an open research task.


\section{The Case for In-Vehicular TSSDN} 
\label{sec:potential_&_expectations}
The OpenFlow standard defines a configuration interface between the controller and the switches in a network.
This enables the vendor neutral selection of controller logic and forwarding devices.
Additionally, the network logic in \ac{SDN} is mostly centralised at the controller.
According to Du et al.\cite{dh-sdnJR-16} these two properties pave the way for simple, exchangeable, inexpensive, and future proof forwarding devices.

The \ac{SDN} controller has global knowledge of the network and all active flows.
This is reinforced by the static nature of the in-vehicular networks.
The network knowledge enables efficient route determination which can prevent link overloading.
On the other hand, it enables the calculation and verification of timings over multiple links during run time.

With the central network knowledge, it is possible to re-route traffic, add new devices (plug \& play) and update the network logic of all switches from a central point.
This simplifies the management of the various configurations of a car model.
For real-time networks this enables the central (re-) calculation and distribution of time-sensitive schedules \cite{ndr-tssJR-16}.

One of the largest potentials of \ac{SDN} in vehicles is adding robustness.
\ac{SDN} supports arbitrary network topologies with redundant paths \cite{dh-sdnJR-16}.
In combination with the global network knowledge, this is an enabler for robust safety methods, such as fail-operational, fail-over or fast re-route through reconfiguration \cite{hmg-rsaJR-18}.
One of the weak points of \ac{SDN} regarding robustness and safety is the single point of failure introduced by a central network controller.
This problem has been solved by using multiple connected controllers \cite{tg-hdcJR-10}, a second controller in hot standby, or a fallback configuration in the switches.

There are already many security suites for \ac{SDN} controllers enabling flow-based firewalls \cite{hhaz-fbrJR-14} or advanced security applications such as anomaly or intrusion detection \cite{vkmca-rtsJR-17}.
To secure in-vehicular networks, the knowledge about the network can be used to only permit statically whitelisted flows in critical sections of the network.

While gaining these advantages and potentials of \ac{SDN} for in-car networks, it must still be possible to maintain the deterministic \ac{QoS} provided by \ac{TSN}.


\section{TSSDN Switching Methodology} 
\label{sec:architecture_proposal}
This section presents the evolution of the switch architecture from standard Ethernet to \ac{TSSDN} switches and describes how OpenFlow can be used to implement time-sensitive flows.

	\subsection{Switch Architecture} 
	\label{sub:switch_archticture}
	\begin{figure}
		\centering
		\includegraphics[width=1.0\linewidth, trim= 1.55cm .5cm 1.55cm .8cm, clip=true]{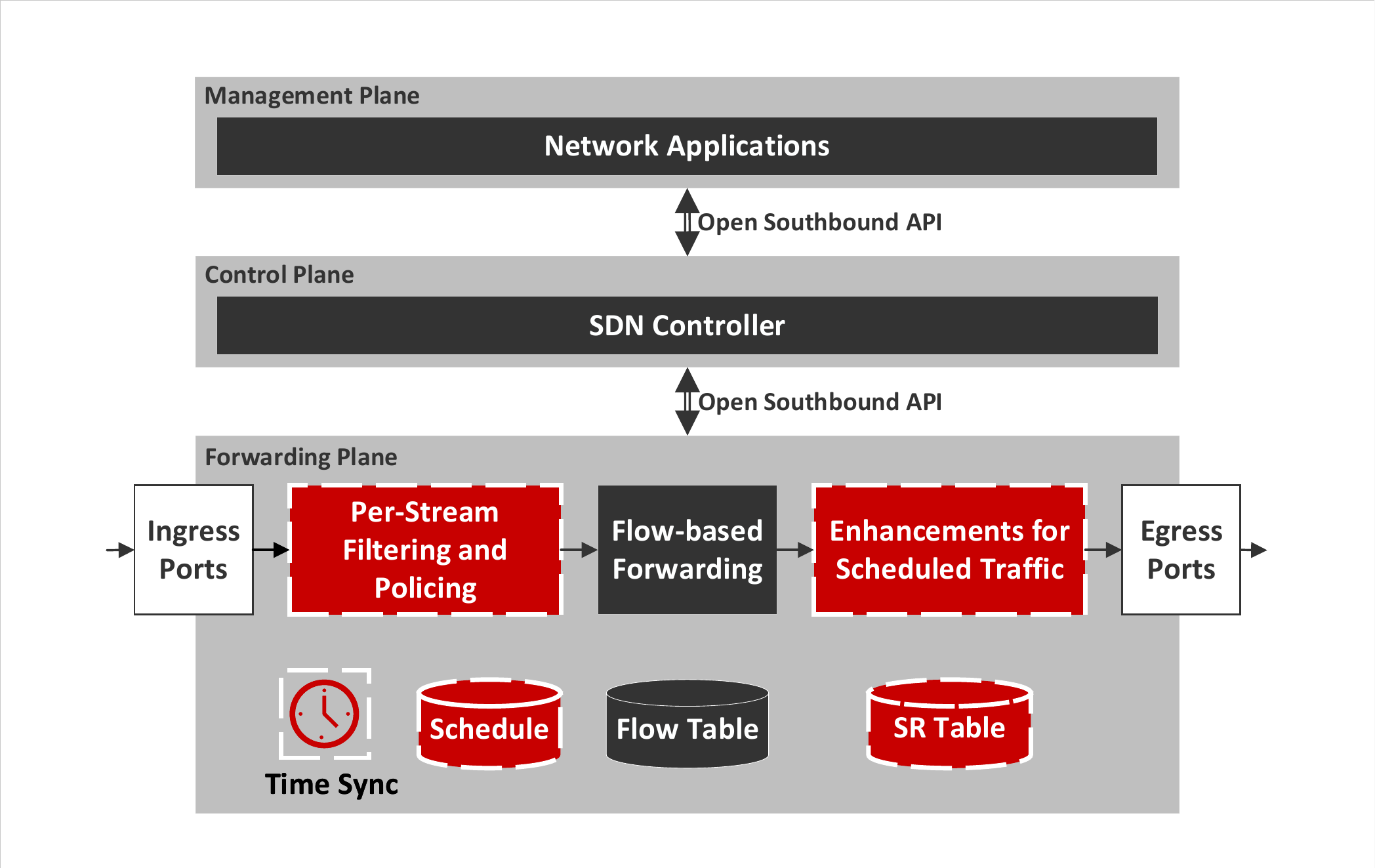}
		\includegraphics[width=0.7\linewidth, trim= .4cm .4cm .4cm .4cm, clip=true]{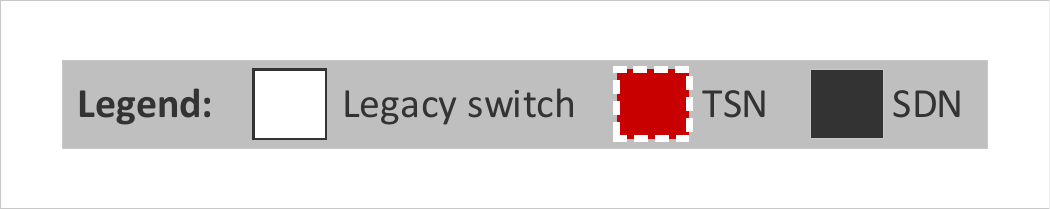}
		\caption{TSSDN Switch Architecture}
		\label{fig:switches}
		\vspace{-10pt}
	\end{figure}

	In \ac{TSN} and \ac{SDN}, switches contain additional modules to extend the functionality of regular switching hardware, which is depicted in Figure~\ref{fig:switches}.
	One of the additional modules introduced for \ac{TSN} is the ``Per-Stream Filtering and Policing'' module.
	It is used to filter incoming Ethernet frames and control the arrival times of \ac{TDMA}-based traffic.
	In \ac{TSN} the egress is controlled by the ``Enhancements for Scheduled Traffic'' module.
	It implements priority queuing, real-time scheduling and traffic shaping.
	The schedule for \ac{TDMA}-based traffic is defined in the ``Schedule'' table and needs a precise and synchronised time in all network devices.
	This is managed by the ``Time Sync'' module.
	The \ac{SRP} is used to dynamically reserve bandwidth along a path of an asynchronous stream.
	The ``SR Table'' module contains all registered talkers and listeners for time-sensitive streams.
	\\
	In \ac{SDN} switches, the forwarding module of a standard Ethernet switch is replaced by a flow-based forwarding module that performs flow table lookups.
	If no corresponding forwarding rule exists, the packet is dropped by default, while most controller applications insert the default rule to forward the packet to the controller.
	Tasks such as topology discovery and route determination are performed by the \ac{SDN} controller.
	The switch is connected to the controller via the Open Southbound API which implements the OpenFlow standard.
	Additional network applications can be executed on top of the controller.

	In \acf{TSSDN}, flow based operations require merging the components of the \ac{TSN} and \ac{SDN} switches.
	To ensure that the real-time capabilities are not altered in any way, the \ac{TSN} ingress and egress control must remain unchanged.
	When data packets arrive, the \ac{TSN} ingress control manages the timing and applies stream-based filters.
	Then the packet is matched against the flow table and the discovered actions in the matching entry are executed.
	The packet is then forwarded to the correct egress ports, where the \ac{TSN} egress control manages the timing and shaping of the outgoing traffic.
	\\
	\ac{TSN} devices exchange additional control information, e.g. to reserve bandwidth with the \ac{SRP}.
	To implement correct flow rules in the forwarding logic, this information needs to traverse to the \ac{SDN} controller.
	The \ac{SDN} controller updates its network state and informs the switch about changes, if needed.
	In this work we implemented a controller application managing the \ac{SR} table for the switches.
	In the future, further parts of the \ac{TSN} control plane could potentially be extracted from the forwarding devices and embedded into the controller.
	Nevertheless, the scheduling and transmission selection needs to remain in the switches to guarantee timing.


	\subsection{Implementing Time-Sensitive Flows with OpenFlow} 
	\label{sub:extending_openflow}
	An extension for \ac{TSSDN} match rules to OpenFlow is needed to control the forwarding of \ac{TSN} flows.
	Still, the additional control information of the \ac{SRP} needs to be communicated between the \ac{TSSDN} switches and the controller.
	\\
	In a normal \ac{TSN} switch, forwarding is performed based on the destination MAC address (multicast) that identifies the stream's listener group.
	By using flow-based forwarding rules in \ac{SDN}, the forwarding can be performed with multiple additional match fields, which enable the realization of non-functional requirements.
	We used the match fields and OpenFlow identifiers shown in Table~\ref{tab:matches} to forward time-sensitive flows.
	By matching the talker source MAC address, we ensure that a stream always originates from the same talker and prevent misuse of the multicast group.
	If the \ac{TSN} path redundancy feature is not in use, a certain \ac{TSN} stream should always arrive at the same ingress port of the switch. 
	If it arrives at another port, the \ac{SDN} controller needs to be informed about changes in the network.
	Other match fields are the VLAN ID and VLAN Stream priority, as a stream multicast group might be used in multiple VLANs.
	This could be useful for future security applications.
	With these match fields, a \ac{TSN} stream can be identified as a flow and can now be forwarded correctly.

	\begin{table}
		\caption{OpenFlow match fields used for \ac{TSN} streams}
		\label{tab:matches}
		\vspace{-2pt}
		\begin{tabular}{l || l | l}
			                 & \textbf{Match field}             & \textbf{OpenFlow Identifier}                \\\hline \hline
			&&\\[-7pt]
			\textbf{Listener Group}		  & Eth. Dst. Addr. 		& OFPXMT\_OFB\_ETH\_DST   \\[0pt]\hline
			&&\\[-7pt]
			\textbf{Talker Address}    	  & Eth. Src. Addr.      	& OFPXMT\_OFB\_ETH\_SRC   \\[0pt]\hline
			&&\\[-7pt]
			\textbf{Ingress Port}          & Switch Ingress Port   	& OFPXMT\_OFB\_IN\_PORT   \\[0pt]\hline
			&&\\[-7pt]
			\textbf{VLAN ID}               & 802.1Q VLAN ID        	& OFPXMT\_OFB\_VLAN\_VID  \\[0pt]\hline
			&&\\[-7pt]
			\textbf{Stream Priority}       & 802.1Q Priority       	& OFPXMT\_OFB\_VLAN\_PCP  \\[0pt]
		\end{tabular}
		\vspace{-14pt}
	\end{table}

	In \ac{TSSDN} all \ac{SRP} messages need to be forwarded to the controller instead of being processed in the switch to enable the controller to configure the forwarding of streams.
	As the OpenFlow protocol does not specify a way to exchange additional control information, we added a new OpenFlow control message type \textit{ForwardSRP}, implementing the standard message signature of OpenFlow and containing the \ac{SRP} Message as a payload.
	In future work this could be mapped to the OpenFlow standard messages.
	\\
	To set up a new stream the talker sends a `talker advertise'.
	The \ac{TSSDN} switch forwards the \ac{SRP} message to the controller.
	The controller then registers the talker in its \ac{SR} table and sends the \ac{SRP} message back to the switch which then updates its own \ac{SR} table and broadcasts the `talker advertise'.
	This way, the talker advertisement is propagated through the network and each switch goes through the same process until the talker advertisement reaches the clients.
	When a client subscribes to a stream by sending a `listener ready' message, the switch forwards it to the controller as well.
	The controller updates the listener in the \ac{SR} table.
	It then pushes a forwarding rule for the \ac{TSN} stream to the switch before it sends the `listener ready' back to the switch.
	When the switch receives the \textit{ForwardSRP} message it delivers it on the direct path to the talker and each switch along the path goes through the same process.
	The talker starts streaming when the `listener ready' arrives.
	Using this approach, we can guarantee the timing since the bandwidth is already reserved and the forwarding rules are already implemented in all forwarding devices along the path.

\section{Case Study} 
\label{sec:case_study}
\begin{figure*}[t]
	\centering
	\begin{minipage}[b]{0.49\textwidth}
	\centering
	\begin{tikzpicture}
		\begin{axis}[width=.95\textwidth, height=.65\textwidth,
	    x unit=\second,
	    change y base,
	    y SI prefix=micro, y unit=\micro\second,
			xlabel=Simulation time,
	    yticklabel pos=left,
			ylabel=End-to-end latency,
	    legend pos=north east,
	    legend cell align={left},
			legend style={font=\small},
	    xmin=0.1,
	    xmax=0.109,
	    xtick distance=0.002,
	    xticklabel style={/pgf/number format/fixed,/pgf/number format/precision=3},
			ymin=0.00009,
	    ymax=0.00155,
	    ]

		\addplot[mark=o,
	    lightcorered,
	    thick,
	    ] table [x, y , col sep=comma] {data/Latency_Vector_AVB_C0toC1_noSDN.csv};
		\addplot[mark=o,
	    coredarkgray,
	    thick,
	    ] table [x, y , col sep=comma] {data/Latency_Vector_UDP_C0toC1_noSDN.csv};
		\addplot[mark=triangle,
	    lightcorered,
	    thick,
	    ] table [x, y , col sep=comma] {data/Latency_Vector_AVB_D0toD1_new.csv};
	  \addplot[mark=triangle,
	    coredarkgray,
	    thick,
	    ] table [x, y , col sep=comma] {data/Latency_Vector_UDP_D0toD1_new.csv};
	  \legend{TSN stream, UDP cross traffic, TSN s. with SDN, UDP ct. with SDN}
		\end{axis}
	\end{tikzpicture}
	\vspace{-8pt}
	\caption{TSN stream and UDP end-to-end latency of the first Ethernet frames}
	\label{fig:startuplatency}
	\end{minipage}
	\hspace{3pt}
	\begin{minipage}[b]{0.49\textwidth}
	  \centering
	\begin{tikzpicture}
	  \begin{axis}[width=.95\textwidth, height=.65\textwidth,
	    ybar,
	    bar width=10pt,
	    change y base,
	    y SI prefix=micro, y unit=\micro\second,
	    enlarge x limits=0.3,
	    legend pos=north west,
	    legend cell align={left},
			legend style={font=\small},
	    yticklabel pos=right,
	    ylabel={End-to-end latency},
	    xtick={0,...,4},
	    xticklabels={, Min, Mean, Max, },
	    x tick label style={align=center},
	    tick align=inside,
			extra y ticks={0.000750},
			extra y tick labels={\textcolor{corered}{750}},
	    xmin=1,
	    xmax=3,
	    ]
	    \addplot[corered,sharp plot,update limits=false] coordinates {(0,0.000750) (4,0.000750)};
			\node[corered] at (axis cs:2.3,0.000820) {TSN};
			\node[corered] at (axis cs:2.3,0.000680) {guarantee};
			\addplot[fill=lightcorered, postaction={pattern color=white, pattern=north west lines}] coordinates {(1,0.000110) (2,0.000390) (3,0.000499)};
	    \addplot[fill=coredarkgray, postaction={pattern color=white, pattern=north west lines}] coordinates {(1,0.000423) (2,0.000481) (3,0.000820)};
			\addplot[fill=lightcorered] coordinates {(1,0.000210) (2,0.000373) (3,0.000483)};
			\addplot[fill=coredarkgray] coordinates {(1,0.000408) (2,0.000466) (3,0.001478)};
	    \legend{, TSN stream, UDP cross traffic, TSN s. with SDN, UDP ct. with SDN}
	  \end{axis}
	\end{tikzpicture}
	\vspace{2pt}
	\caption{TSN stream and UDP min/max end-to-end latency of Ethernet frames}
	\label{fig:minmaxlatency}
	\end{minipage}
	\vspace{-28pt}
\end{figure*}

Our case study analyses the real-time capabilities of the proposed \ac{TSSDN} concept in a simulation environment.
Our simulation is based on the open-source OMNeT++ IDE and the INET framework \cite{omnetpp-inet-framework}.
It provides simulation models of standard Internet technologies. 
On top, we use the CoRE4INET framework\cite{sdkks-eifre-11}\cite{core-frameworks} which implements real-time Ethernet protocols and the OpenFlowOMNeTSuite\cite{g-oJR-16}\cite{kj-oeoJR-13} which implements the OpenFlow protocol.
For the proposed \ac{TSSDN} concept, we merged the CoRE4INET framework and the OpenFlowOMNeTSuite in parts.

\begin{figure}
  \centering
	\includegraphics[width=\linewidth, trim=17pt 17pt 17pt 17pt, clip=true]{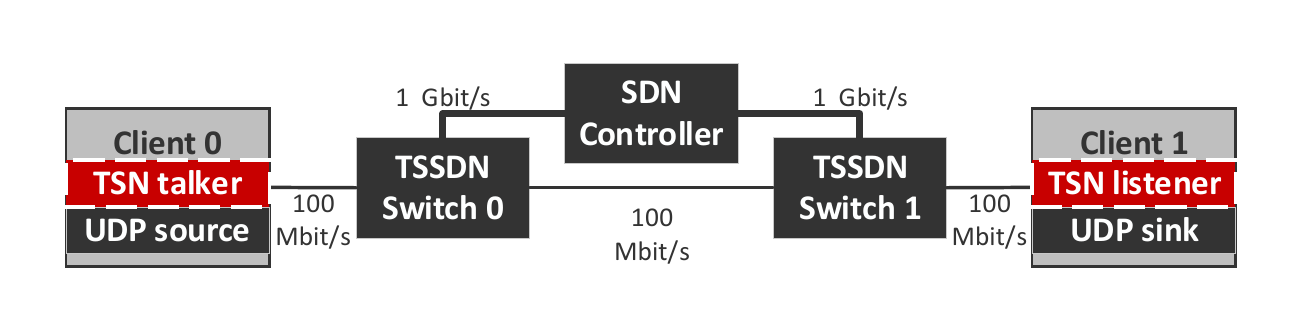}
	\caption{Network topology of the case study}
	\label{fig:case_study_topology}
	\vspace{-17pt}
\end{figure}

The network topology of our simulation-based case study consists of two clients, two \ac{TSSDN} switches, and one SDN controller as shown in Figure~\ref{fig:case_study_topology}.
Client~0 is the source for the TSN stream and the UDP cross traffic.
Client~1 is the receiver of all traffic.
With this setup, the timing requirements of \ac{TSN} across multiple links under load of cross traffic can be analysed.
For comparison, the identical scenario without \ac{SDN} using \ac{TSN}-only switches was considered.
All flows remain the same, so that the result can be directly compared.
\\
At the beginning of the simulation, all the flow tables in the \ac{TSSDN} switches are empty. 
After the controller and switches are connected to each other as specified in the OpenFlow standard, the controller transmits an updated default action for table misses, to be forwarded to the controller for further decision.
For this reason, an idle time of \SI{100}{\ms} for setup operations is introduced in both versions of the network.
After that, the clients can start to work.

Figure~\ref{fig:startuplatency} shows the latency during the connection setup phase for all frames transferred in \SI{9}{\ms} after the start of transmission.
It also compares the latency of the networks with and without \ac{SDN}.
\\
To set up the \ac{TSN} stream, the talker advertises the stream and its listeners subscribe with the help of the \ac{SRP}.
In the \ac{TSSDN} version of the network, the controller receives the \ac{SRP} messages from the forwarding devices, updates its \ac{SR} tables and implements the correct flow entries in the forwarding devices.
In our simulation this introduces a delay of \SI{0.3}{\ms} to the \ac{SR} process compared to the network without SDN.
When the talker receives the listener ready, it begins to stream.
As the stream is already registered in the \ac{TSSDN} switches during the \ac{SR}, no further inspection by the \ac{SDN} controller is needed.
Thus, no additional latency is introduced by \ac{SDN} after the \ac{SR}.
\\
Before the UDP source starts sending frames, it resolves the cross traffic destination MAC address by ARP.
In the \ac{TSSDN} version of the network those ARP frames are forwarded by the \ac{SDN} controller, which again introduces a slight delay.
Thereafter, the UDP source starts sending datagrams.
There is no matching flow entry for the first UDP frame in the \ac{TSSDN} switches.
UDP frames are sent to the controller and the flow tables are updated accordingly.
This leads to a much higher latency for the first UDP frames than for the \ac{TSN} stream.
\\
In the \ac{TSSDN} version of the network, the latency of the \ac{TSN} stream increases slowly as the cross traffic traverses the links.
At first, only the link between client~0 and switch~0 is under load of cross traffic.
When the controller implements the flow rule in switch~0, the UDP frames put load on the link between switch~0 and 1.
The latency for the \ac{TSN} stream increases step by step, until all network links are under load of cross traffic.
In the version without \ac{SDN}, the UDP cross traffic effects the \ac{TSN} stream immediately. We examined the impact of cross traffic on time-sensitive communication in previous work\cite{slksh-bhcan-15}.
\\
Due to the delay of the first UDP frames in the \ac{TSSDN} version of the network, the UDP cross traffic builds up at the \ac{TSSDN} switches which increases the latency. The latency normalises after \SI{7}{\ms}. Afterwards both networks behave identical.

Figure~\ref{fig:minmaxlatency} shows the minimum, maximum and average latency for the transmission of Ethernet frames from client~0 to client~1 and compares the simulation results of the UDP cross traffic and the \ac{TSN} stream for both versions of the network.
The guarantee of \SI{750}{\us} shows the analytic maximum latency for the \ac{TSN} stream.
As the \ac{TSN} stream in this example has the highest possible priority, the maximum latency introduced per port scheduling process is \SI{250}{\us} (see max. latency AVB Stream Class A \cite{ieee8021ab}).
The \ac{TSN} stream is scheduled at three output ports along the path: {Client 0}, {Switch 0} and {Switch 1}.
\\
The minimum latency of the \ac{TSN} stream is lower than that of the UDP cross-traffic in both versions of the network, because the transmission of cross traffic starts with some delay.
After the initialization of the UDP path, the minimum latency is \SI{320}{\micro\second} for the \ac{TSN} stream and \SI{100}{\micro\second} higher for UDP in both networks.
The average latency is very similar for all transmissions.
This is expected as \ac{TSN} only guarantees a maximum latency for the \ac{TSN} stream.
The maximum latency of the UDP frames in the \ac{TSSDN} version of the network is a result of the delay introduced by the \ac{SDN} controller implementing the flow rules.
After the setup phase, the maximum Latency for UDP is \SI{910}{\micro\second} which is still above the \ac{TSN} guarantee.
In the version of the network without \ac{SDN} the maximum latency for UDP is about \SI{820}{\micro\second}.
Figure~\ref{fig:minmaxlatency} shows that the time-sensitive streams are not delayed by \ac{SDN} control traffic and all deadlines are met. The minimum, maximum and average latency of the \ac{TSN} streams is about the same for both versions of the network.

Our results show that the combination of the \ac{SDN} paradigm and \ac{TSN} in the proposed architecture works as expected.
The real-time traffic deadlines for the \ac{TSN} streams are met and not affected by the introduction of \ac{SDN}.
At the same time, the forwarding is controlled by an \ac{SDN} controller which enables the potentials described in Section~\ref{sec:potential_&_expectations}.


\section{Conclusion \& Outlook} 
\label{sec:conclusion_&_future_work}
This paper made the case for in-vehicular \acl{TSSDN} and explored its potentials and expectations. 
We presented our proposal of an integrated \ac{TSSDN} switching that combines \ac{TSN} and \ac{SDN} capabilities. 
The core of this approach is an implementation of time-sensitive flows via OpenFlow, and we specified in detail how \ac{TSN} streams can be mapped and registered at SDN controllers. 
In a case study of \ac{TSSDN} in an event-based simulation environment, we could demonstrate that while gaining all the potentials of \ac{SDN} in the automotive network its real-time capabilities remain unaltered.

In future work, we plan to investigate possibilities of transferring further parts of the \ac{TSN} control logic into the \ac{SDN} controller. 
One important part will be the implementation of static \ac{TDMA} flows in a \ac{TSSDN} and the analysis of possible side effects of \ac{SDN} on such time triggered flows. 
Another open research challenge will be to investigate real-world potentials for vehicles including desired improvements of robustness and security. 
In addition, it will be of interest whether those security mechanisms could also be applied to time-sensitive flows without sacrificing the timing constraints.

\section*{Acknowledgements}
This work is funded by the Federal Ministry of Education and Research of Germany (BMBF) within the SecVI project.


\bibliographystyle{IEEEtran}
\bibliography{bib/all_generated,bib/bibliography}

\end{document}